\documentclass[fleqn,usenatbib]{mnras}

\usepackage{newtxtext,newtxmath}

\usepackage[T1]{fontenc}

\DeclareRobustCommand{\VAN}[3]{#2}
\let\VANthebibliography\thebibliography
\def\thebibliography{\DeclareRobustCommand{\VAN}[3]{##3}\VANthebibliography}

\usepackage{graphicx}	
\usepackage{amsmath}	
\usepackage{xcolor}

\title[The influence of galaxies on their local IGM]
{Exploring the influence of reionisation-era galaxies on their local intergalactic medium using the Sherwood-Relics simulations}

\author[H. Anderson et al.]{Hamish Anderson$^{1}$\thanks{E-mail: handers@mpa-garching.mpg.de},
Laura C. Keating$^{2}$,
Luke Conaboy$^{3}$,
Xiangyu Jin$^{4}$, 
James S. Bolton$^{3}$ 
\newauthor
and Ewald Puchwein$^{5}$
\\
$^{1}$Max Planck Institut für Astrophysik, Karl Schwarzschild Straße 1, D-85741 Garching, Germany\\
$^{2}$Institute for Astronomy, University of Edinburgh, Blackford Hill, Edinburgh, EH9 3HJ, UK\\
$^{3}$School of Physics and Astronomy, The University of Nottingham, University Park, Nottingham, NG7 2RD, UK\\
$^{4}$Department of Astronomy, University of Michigan, 1085 S. University Ave., Ann Arbor, MI 48109, USA\\
$^{5}$Leibniz-Institut für Astrophysik Potsdam, An der Sternwarte 16, D-14482 Potsdam, Germany\\
}

\date{Accepted XXX. Received YYY; in original form ZZZ}

\pubyear{\the\year{}}

\begin{document}
\label{firstpage}
\pagerange{\pageref{firstpage}--\pageref{lastpage}}
\maketitle

\begin{abstract}
Recent surveys have detected excess flux in the Lyman-$\alpha$ forest of quasar spectra several Mpc from high-redshift line-emitting galaxies, indicating a correlation between intergalactic medium (IGM) transmission and reionisation-era galaxies. We examine the galaxy-IGM connection using the Sherwood-Relics simulation suite, a set of hybrid radiation-hydrodynamic simulations that model the IGM at high redshift. We simulate the Lyman-$\alpha$ forest and a population of mock [\ion{O}{iii}] emitters using the survey specifications of JWST's ASPIRE programme. We measure the evolution of the effective optical depth in regions close to galaxies, and contrast this with a control sample probing the random IGM. We characterise the changing physical properties of the IGM to explain the observed sightline-to-sightline scatter in opacity between $5.4<z<6.2$. We find that the IGM is more opaque than average at small radii ($r\leq$ 5 cMpc/$h$) close to galaxies, as these regions are overdense. At larger radii ($15 < r \leq 50~{\rm cMpc}/h$), the IGM is more ionised than average, due to the higher photoionisation rate near galaxies. We find qualitative agreement with the observational results from ASPIRE, though we see large realisation-to-realisation scatter driven by cosmic variance, as also seen in theoretical explorations of the galaxy–Ly$\alpha$ transmission cross-correlation function. Our results generally agree with observations claiming that there is a large-scale ionisation bias around [\ion{O}{III}] emitters, supporting the idea that their role is significant in shaping the properties of the IGM during reionisation.
\end{abstract}




\section{Introduction}

The Epoch of Reionisation (EoR) was the last major phase transition of the Universe, where UV photons produced by the first luminous sources ionised the previously neutral intergalactic medium (IGM). Understanding this transition can reveal the properties of early galaxy formation. However, fundamental questions remain regarding the nature of the sources responsible for driving this process, in large part due to the difficulty in measuring the escape fraction of ionising photons from galaxies \citep{jaskot2025}. Whether the escape fraction is higher in lower-mass or higher-mass galaxies will alter the timing of reionisation \citep{Robertson2015, Finkelstein2019, Naidu2020} as well as the morphology of ionised bubbles \citep{cain2023, lu2024}. These escape fractions cannot be measured directly during reionisation due to the high opacity of the IGM \citep{Inoue2014}. Instead, they are typically inferred from correlations with other galaxy properties measured from surveys of lower-redshift Lyman-continuum emitting galaxies \citep[e.g.,][]{jaskot2024,Mascia2024}.

Alternatively, the escape fraction of ionising photons can be constrained by characterising how galaxies modulate the ionisation state of the local IGM. The Lyman-$\alpha$ (Ly$\alpha$) forest in the spectra of high-$z$ quasars serves as an effective tracer for this, as the Ly$\alpha$ optical depth is sensitive to small changes in the \ion{H}{i} fraction \citep[e.g.,][]{McQuinn2016}. Correlating galaxy positions with Ly$\alpha$ transmission can therefore reveal how these sources ionise their surroundings. Post-reionisation, this cross-correlation shows increased absorption near galaxies \citep{Adelberger2005, Turner2014, Bielby2017, Matthee2024, Banerjee_2025}. These regions are opaque to Ly$\alpha$ photons due to surrounding overdensities and self-shielded residual \ion{H}{i}. Approaching higher redshifts ($z\gtrsim5-6$), the global ultraviolet background (UVB) decreases due to a reduction in the mean free path of ionising photons \citep{Becker2021, Zhu_2023, gaikwad2023, Davies2024}. In this case, when there is a background quasar sightline at a small impact parameter from a galaxy, the quasar Ly$\alpha$ forest can exhibit excess IGM transmission relative to the cosmic mean, if there is a significant local contribution to the ionising background from the galaxy itself. Indeed, early ground-based observations have probed this regime by measuring the Ly$\alpha$ forest near reionisation-era galaxies and metal absorption systems \citep{Kakiichi2018, Meyer2019, Meyer2020}. The resultant galaxy-Ly$\alpha$ cross-correlation function has revealed both excess absorption close to galaxies ($\lesssim 5\;\mathrm{cMpc}/h$) and enhanced Ly$\alpha$ transmission at intermediate distances ($\gtrsim 10\;\mathrm{cMpc}/h$). As the UVB diminishes, the local radiation field of galaxies begins to dominate, supporting the consensus that clustered, faint sources provide an important contribution to the reionisation budget \citep{Kakiichi2018, Meyer2019}. The statistical significance of these trends remains uncertain, reflecting the observational limits of current ground-based surveys \citep{Meyer2020}. 

The high sensitivity and infrared capabilities of JWST have made it easier to obtain simultaneous line-of-sight spectroscopy and detection of galaxies with rest-optical lines like [\ion{O}{iii}], which, unlike Ly$\alpha$, are unaffected by IGM attenuation. Surveys in quasar fields allow for correlation with deep ground-based Ly$\alpha$ forest data \citep{Kashino_2023, Wang_2023}. New analyses of the Ly$\alpha$ forest-galaxy cross-correlation at $z\sim 6$ \citep{Kashino_2023, Kakiichi2025, Kashino2026, Jin2026, Zhu2026} have confirmed that an excess of Ly$\alpha$ forest transmission in the spectra of high-redshift quasars is observed in the vicinity of line-emitting galaxies, which is in qualitative agreement with previous ground-based results \citep{Kakiichi2018, Meyer2019}. However, the exact strength of the peak in the cross-correlation function and the distance at which it occurs among the different surveys are still somewhat uncertain, complicated by a combination of cosmic variance, differing galaxy detection methods and varying survey resolutions. The relative roles of galaxies and AGN in setting the amplitude and scale of the cross-correlation also remain uncertain \citep{Jin2026}.

Cosmological simulations that model an inhomogeneous reionisation are found to recover broadly consistent results for the galaxy-Ly$\alpha$ correlation, where Ly$\alpha$ transmission is suppressed close to galaxies ($r\lesssim10$ cMpc/$h$) and enhanced at further distances ($r\gtrsim10$ cMpc/$h$) \citep{Garaldi2022, Garaldi2024, Conaboy2025, Basu2025}. The cross-correlation is governed by two competing effects: overdense regions closest to galaxies support recombination and \ion{H}{i} self-shielding, whereas ionising photon density dominates at further distances to boost Ly$\alpha$ transmission \citep{Garaldi2024}. Additionally, the predicted amplitude of excess transmission is found to be sensitive to the assumed spectral energy distribution of the ionising sources \citep{Basu2025}. The source population model and the applied survey geometry tend to have a smaller impact on the shape of the correlation curve, but significantly influence the scatter \citep{Garaldi2024, Conaboy2026}. Conversely, other models of absorption spectra have reported no excess transmission at large distances \citep{Garaldi2019, Zhu2024b}, suggesting sensitivity to the galaxy formation model, timing of reionisation or radiative transfer implementation.

Beyond the cross-correlation function, measuring the effective optical depth of the Ly$\alpha$ forest is a highly sensitive probe of the tail-end of reionisation \citep{Fan2006, Becker2015, Bosman2022}. The effective optical depth $\tau_\text{eff}$ is defined as $\tau_\text{eff}=-\ln\langle F\rangle$, where $\langle F\rangle$ is the mean flux of the Ly$\alpha$ forest measured over a fixed distance, typically 50 cMpc/$h$. The method outlined in \cite{Jin2024} presents an alternative way to measure correlation between galaxies and the IGM, linked to these traditional studies of the Ly$\alpha$ forest, by measuring the distribution of Ly$\alpha$ forest opacities near [\ion{O}{iii}] emitters. They found that, when computed over a sufficiently large path length ($\gtrsim50 \;\mathrm{cMpc}/h$), the optical depth close to galaxies was consistently lower compared to that of the more distant IGM. This reinforces the role of clustered galaxies in driving late-stage reionisation and indicates the observed scatter in $\tau_{\mathrm{eff}}$ is tightly coupled to fluctuations in the UVB \citep{Jin2024}. 

This work aims to critically evaluate the observational trends presented in \cite{Jin2024} using one of the Sherwood-Relics suite of hybrid radiation-hydrodynamical simulations \citep{Puchwein2022}. In Section \ref{sec:2}, we describe these simulations and how we construct a mock observational survey that follows the specifications of the ASPIRE programme, to generate a distribution of Ly$\alpha$ $\tau_\text{eff}$ measurements. In Section \ref{sec:3}, we compare our models with the observations and explore how these measurements evolve with redshift, as well as their dependence on assumptions about the underlying galaxy population. In Section \ref{sec:4}, we explore the gas properties of the evolving IGM, to determine the physical drivers of the emerging trends in $\tau_{\text{eff}}$. In Section \ref{sec:5}, we present our conclusions.

Throughout this work, we assume a flat $\Lambda\text{CDM}$ cosmology consistent with the results from the \cite{Planck2014}, with $\Omega_m = 0.308$, $\Omega_\Lambda = 0.692$, $\Omega_b = 0.0482$, and $h = 0.678$. Comoving and proper distances are represented by the prefixes c and p respectively.

\begin{figure*}
    \centering
    \begin{minipage}{\textwidth}
        \centering
        \includegraphics[width=\linewidth]{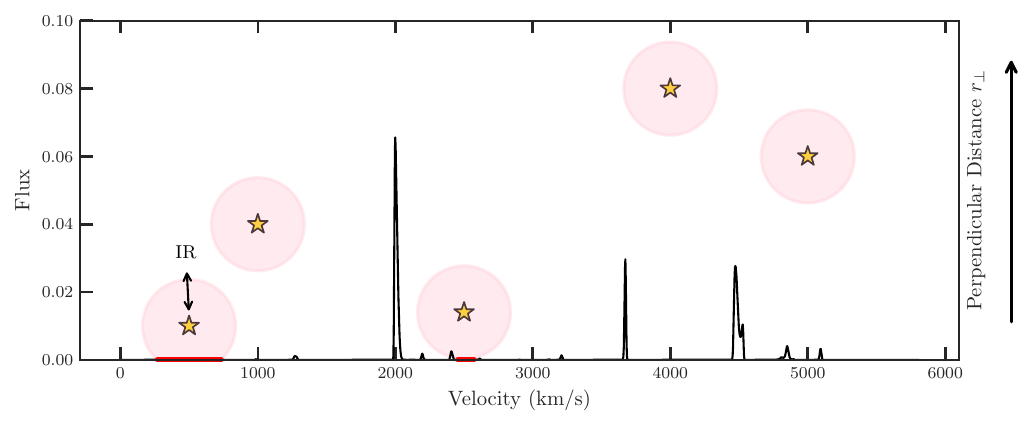}
    \end{minipage}
    
    \vspace{0.4cm}

    \begin{minipage}{0.48\textwidth}
        \vspace{0.6cm}
        \centering
        \includegraphics[width=\linewidth]
        {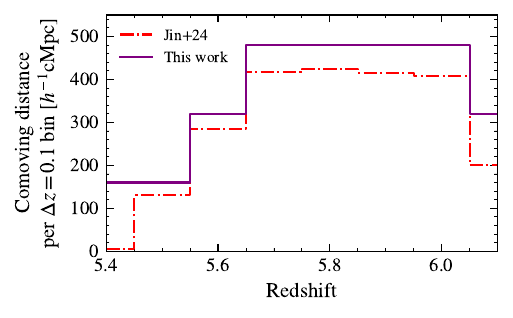}
    \end{minipage}
    \hfill
    \begin{minipage}{0.48\textwidth}
        \centering
        \includegraphics[width=\linewidth]{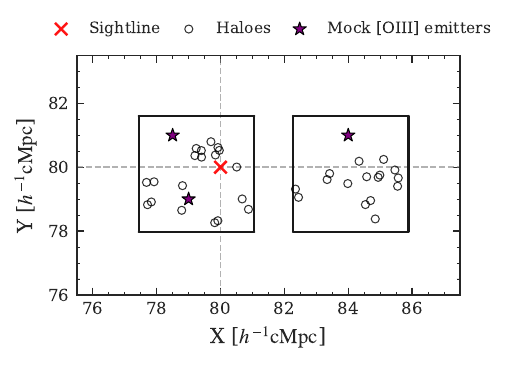}
    \end{minipage}
 
    \caption{\textbf{Top}: Visualisation of the sampling method used to compute source-biased $\tau_{\mathrm{eff}}$ across mock quasar spectra. The black line shows an example of a mock Ly$\alpha$ forest spectrum from our simulation and the yellow stars are mock [\ion{O}{iii}] emitters. A sphere of influence with an associated influence radius (IR) is placed around nearby [\ion{O}{iii}] emitters. The mean flux is computed over distances along the sightline determined by the path length enclosed by the intersecting influence sphere, shown by the thick red line. The resulting $\tau_{\mathrm{eff}}$ distribution is therefore sensitive to the positions of nearby sources. The relative scales and represented number of [\ion{O}{iii}] emitters are purely illustrative. \textbf{Bottom Left}: Summary of the combination of sightlines we use to construct our mock survey as a function of redshift. The red dash-dotted line shows the comoving distance probed by the \citet{Jin2024} analysis of the ASPIRE survey measured in intervals of $\Delta z=0.1$, corresponding to the redshift spacing of our line-of-sight outputs. In purple, we show the comoving distance covered by our mock survey. Note that this does not always agree with the observed path length, as we are forced to use an integer number of sightlines through our simulation in order to make use of periodic boundary conditions. \textbf{Bottom Right}: Illustration of a mock survey that replicates the dimensions and geometries of NIRCam pointings. The open circles are the location of haloes in our simulation and the purple stars show the sources that have been randomly selected to be [\ion{O}{iii}] emitters. The red cross corresponds to the position of the quasar, as placed in the ASPIRE pointings. Two symmetric detectors of width 3.62 cMpc/$h$ define the area within which [\ion{O}{iii}] emitters are selected. Sightlines are placed at the same offset relative to the centre of the FOV, which itself is shifted to the centre of the box to simplify boundary conditions. The search radius for [\ion{O}{iii}] emitters is restricted by the boundaries imposed by the FOV.}
    \label{fig:intro}
\end{figure*}

\section{Modelling the connection between galaxies and the Ly$\alpha$ forest}
\label{sec:2}

\subsection{Reionisation simulation}
\label{sec:2.1}

We analyse one of the patchy reionisation simulations from the Sherwood-Relics simulation suite \citep{Puchwein2022}, which builds upon the original Sherwood suite \citep{bolton2017}. To sample a wide range of environments for simulated [\ion{O}{iii}] emitters, we look at a 160 cMpc/$h$ volume with 2048$^3$ gas particles performed with the cosmological hydrodynamic code \textsc{p-gadget3} (last described in \citealt{Springel2005}). Star formation in the simulation is treated in a simplified way, turning all gas with density more than 1000 times the cosmic mean and with temperatures less than 10$^5$ K immediately into star particles. It has been shown that this does not change the properties of the high-redshift IGM \citep{Viel2004}. This patchy simulation utilises a novel treatment of reionisation, described in detail in \cite{Puchwein2022}, whereby time-evolving 3D maps of the UV background are generated by post-processing a simulation of the same volume and mass resolution with the \textsc{aton} radiative transfer code  \citep{Aubert2007}. As we do not use a self-consistent galaxy formation model, we place ionising sources at locations of haloes with masses $M_{\rm h} > 10^9 \; M_{\odot}/h$ and assign these a luminosity proportional to the halo mass. A new ``hybrid'' simulation is then performed with these UV background maps, to account for inhomogeneous photoionisation and photoheating of the gas. In this particular simulation, reionisation ends at $z=5.3$, in agreement with data from the Ly$\alpha$ forest \citep{Bosman2022}. It is important to note that our simple source modelling will not produce a realistic population of [\ion{O}{III}] emitters, where line emission is sensitive to a range of physical processes within the ISM \citep{Casavecchia2026} and which can produce significant scatter in ionising photon production and escape at fixed halo mass.

To model the Ly$\alpha$ forest, we analyse 5000 periodic lines of sight extracted from the simulation, with pixel size $\Delta x = 78.11\; \text{ckpc}/h$. From the gas properties at each pixel, we calculate the optical depth using the analytic Voigt profile approximation from \cite{Garcia2006} to construct Ly$\alpha$ spectra in velocity space. We note that we correlate these spectra with halo positions in real space, introducing an inconsistency into our analysis. However, peculiar velocities are only expected to have a small effect on galaxy-IGM cross-correlation statistics \citep{Meyer2020,Garaldi2024} and should not change any of our conclusions.

\subsection{Measuring the IGM effective optical depth near galaxies}
\label{sec:2.2}

\begin{figure*}
    \centering
    \begin{minipage}[b]{0.48\textwidth}
        \centering
        \includegraphics[width=\linewidth]{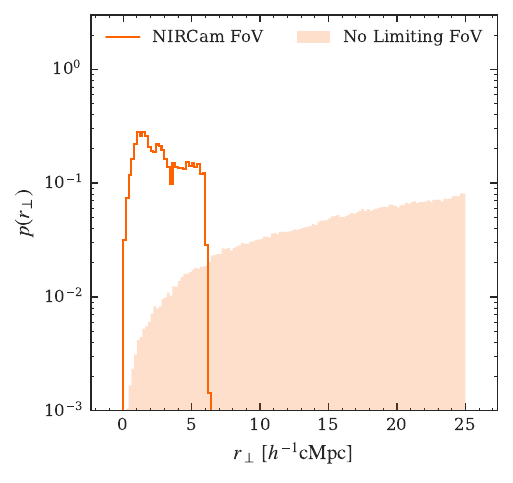}
    \end{minipage}
    \hfill
    \begin{minipage}[b]{0.48\textwidth}
        \centering
        \includegraphics[width=\linewidth]{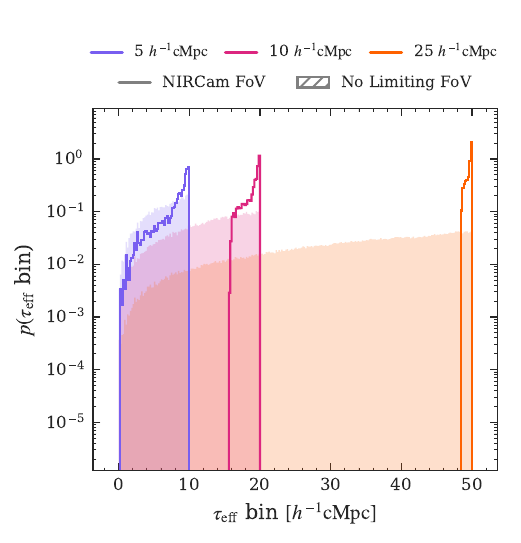}
    \end{minipage}
    \caption{\textbf{Left}: The probability distribution of the perpendicular distances between a sightline and surrounding sources for an influence radius of 25 $\mathrm{cMpc}/h$, before and after imposing a limiting FOV. Without restrictions, the range of [\ion{O}{iii}] emitter distances probed is determined by the influence radius itself, highlighted by the shaded orange distribution. With an imposed FOV, the dimensions of the NIRCam module set the maximum search radius ($r_{\mathrm{max}} = 5.83 \; \mathrm{cMpc}/h$), illustrated by the solid orange line. \textbf{Right}: The distribution of $\tau_{\mathrm{eff}}$ bin sizes determined by influence spheres about a sightline, before and after imposing a limiting FOV. For small influence radii (IR = 5 $\mathrm{cMpc}/h$ in blue), the bin sizes to compute $\tau_{\mathrm{eff}}$ show a distribution spanning from 0 to 2 times the influence radius. Once the influence radius exceeds the size of the NIRCam FOV, the width of the distribution decreases, e.g., for the largest radius shown here (IR = 25 $\mathrm{cMpc}/h$ in orange), the bin size is roughly two times the influence radius for all selected [\ion{O}{iii}] emitters. The background shaded regions highlight the range of $\tau_{\mathrm{eff}}$ bin sizes when no FOV restrictions are used.}
    \label{fig:nircam_distributions}
\end{figure*}

\subsubsection{Spheres of Influence}
\label{sec:2.2.1}

In order to determine the influence of ionising sources on surrounding mock sightlines, we reproduce the galaxy-sampling method outlined in \cite{Jin2024}. This is summarised in Fig. \ref{fig:intro} (top). For a given sightline, we place spheres with a specified influence radius (IR) around nearby [\ion{O}{iii}] emitters. The spheres that intersect the sightline determine the  [\ion{O}{iii}] emitters which are close enough to influence the opacity of the corresponding Ly$\alpha$ forest. Moreover, the bin size used to calculate $\tau_{\mathrm{eff}}$ along the sightline is determined by the enclosed path length of the intersecting sphere, indicated by the thick red line. Source-sightline pairs are treated independently, i.e., two overlapping influence spheres produce two individual measurements. For small IR, this also ensures that [\ion{O}{iii}] emitters closer to a sightline contribute more to the resulting $\tau_{\mathrm{eff}}$ distribution for a particular spectrum. Therefore, we can construct a large sample of $\tau_{\mathrm{eff}}$ measurements which are coupled to the relative positions of nearby ionising sources, effectively simulating the galaxy proximity effect observed in reionisation-era surveys. The advantage of this approach is that it can be applied to quasar spectra with varying spectral resolution and signal-to-noise, which may not be suitable for calculations of galaxy–Ly$\alpha$ transmission cross correlation function \citep{Kakiichi2025}. On average, a short influence radius traces the overdense IGM near [\ion{O}{iii}] emitters, while a larger influence radius will account for more distant structures. For comparison, we construct a second sample of $\tau_{\mathrm{eff}}$ measurements to evaluate the average IGM over varying length scales. Here, we draw another set of sightlines and reuse the same path lengths to compute a new $\tau_{\mathrm{eff}}$ distribution, while ignoring the relative positions of [\ion{O}{iii}] emitters. With enough repeats, we expect this sample to trace the average IGM. We find that the influence radius serves as an observationally motivated diagnostic that not only probes radial distance, like conventional cross-correlation studies, but simultaneously scales the $\tau_{\mathrm{eff}}$ bin size, thereby encoding characteristic scale-length information. We discuss below how this effect is convolved with the survey footprint.

\subsubsection{Mock Survey Criteria}
\label{sec:2.2.3}

It is important to consider the number of sightlines required to make a fair comparison between the Sherwood-Relics simulation and the observational data from the ASPIRE survey. The survey measures $\tau_{\mathrm{eff}}$ of the Ly$\alpha$ forest across 14 quasar sightlines. They analyse the Ly$\alpha$ forest between rest-frame wavelengths 1040 \AA \, and 1176 \AA, beyond which contamination from Ly$\beta$ and the quasar proximity zone respectively become prominent. The ASPIRE quasars that were analysed range from $z\sim6.5$ to $z\sim6.8$, and the total Ly$\alpha$ forest spans from $z\sim5.4$ to $z\sim6.6$ \citep{Jin2024}, however only data up to $z\sim6.1$ is considered in their analysis, since beyond this only lower limits to the Ly$\alpha$ forest opacity are measured. To capture the rapid evolution of the IGM towards the end-stages of reionisation \citep[e.g.,][]{keating2020}, we create a single mock survey from outputs from our simulation at different redshifts, spaced at $\Delta z = 0.1$ intervals. We demonstrate why this is necessary, as opposed to using the redshift at the midpoint of the survey, in Section \ref{sec:3.3}. We show in the bottom left panel of Fig. \ref{fig:intro} how the path length of the analysis in \citet{Jin2024} corresponds to our mock survey. We limit ourselves to integer numbers of sightlines such that influence spheres can be treated periodically, hence the ASPIRE path length is not matched exactly. We found that truncating sightlines discards some nearby haloes from the sample which would otherwise influence the Ly$\alpha$ forest, partially breaking the correlation between halo position and ionisation state. This occurs naturally at the edges of the observed Ly$\alpha$ forest, but since a single sightline through our volume is shorter than the Ly$\alpha$ forest probed by a single ASPIRE quasar, we predict that truncating sightlines to match the redshift distribution of observed path lengths can introduce extra uncertainty into our analysis. We repeat these mock surveys 100 times to determine the effects of cosmic variance among our realisations.

Alongside the mock Ly$\alpha$ forest data, we also construct a population of mock [\ion{O}{iii}] emitters, as we do not model these directly in our simulation. Unlike in \citet{Conaboy2026}, we do not derive these from abundance matching but rather make use of existing constraints on the halo masses of [\ion{O}{iii}] emitters \citep{Eilers2018, Huang2026}. Based on such analyses, we choose a minimum host halo mass for our mock [\ion{O}{iii}] emitters of $\log(M_{\text{halo}}/M_\odot)=10.55$ and a duty cycle of $2.5\%$. We use these parameters to randomly select a population of [\ion{O}{iii}] emitters from the haloes in our simulation. An example of this is shown in the bottom right panel of Fig. \ref{fig:intro}. Only a small fraction of the haloes in our simulation (open circles) are identified as [\ion{O}{iii}] emitters (purple stars). Later, we also examine the impact of changing the minimum halo mass (section \ref{sec:3.2}). In this case, we increase or decrease the duty cycle accordingly such that the number density of sampled haloes remains constant.

\subsubsection{NIRCam Field of View Effect}
\label{sec:2.2.2}

The source sampling in this paper adheres to the out-of-field imaging limits of the ASPIRE programme \citep{Wang_2023}, which employs single NIRCam pointings. We replicate the survey geometry (Fig. \ref{fig:intro}), consisting of two 3.62 $\times$ 3.62 cMpc/$h$ (2.2' $\times$ 2.2') modules separated by a 1.2 cMpc/$h$ (44'') gap \citep{rieke2023}, as shown in the bottom right panel of Fig. \ref{fig:intro}. We place mock sightlines at the same relative offset within Module A (left) as the ASPIRE quasars: $X= -1.66$ cMpc$/h$ (-60''.5) and $Y=0.21$ cMpc$/h$ (7''.5) \citep{Wang_2023}. For convenience, the sightlines and surrounding mock [\ion{O}{iii}] emitters are shifted to the centre of the box to simplify the sampling procedure for sightlines near the box boundary. 

The effect of imposing NIRCam's field of view (FOV) has important consequences on the length scales probed by a given influence radius. Firstly, the FOV truncates the distribution of perpendicular distances between a source and a sightline. With no FOV restrictions, the influence radius typically probes a continuous distribution of perpendicular distances, the maximum of which is set by the influence radius itself, illustrated by the orange shaded distribution in Fig. \ref{fig:nircam_distributions} (left). In this case, the total number of [\ion{O}{iii}] emitters probed near a given sightline increases with influence radius. With the restrictions imposed by the FOV, the maximum perpendicular distance is set by the dimensions of the detector modules ($r_{\text{max}} = 5.83 \; \mathrm{cMpc}/h$), highlighted by the solid orange curve. The offset of the sightline within module A introduces an asymmetric search radius, indicated by the asymmetry of the distribution, with more [\ion{O}{iii}] detected at small radii. Additionally, for IR $<r_{\text{max}}$, not all  mock [\ion{O}{iii}] emitters  within the FOV are close enough to the sightline to be included in the analysis. For IR $>r_{\text{max}}$, the total number of sampled [\ion{O}{iii}] emitters remains the same as we probe the maximum number of [\ion{O}{iii}] emitters within a given pointing.

Increasing the influence radius beyond this limit purely affects the line-of-sight distance over which the mean flux of the Ly$\alpha$ forest is measured. This is demonstrated in the right panel of Fig. \ref{fig:nircam_distributions}. For IR $>r_{\text{max}}$, the bin size used to compute $\tau_\text{eff}$ is typically twice the influence radius. For IR $<r_{\text{max}}$, the NIRCam restriction has a smaller effect on $\tau_\text{eff}$ as these influence spheres fit entirely within the FOV of the pointing. In this case, the $\tau_\text{eff}$ bin size ranges from zero to twice the influence radius. Therefore, short influence radii probe a range of length scales across a given sightline, whereas larger radii probe a similar bin size for all mock [\ion{O}{iii}] emitters regardless of their relative positions within the pointing. This means that in most cases, IR = 25 cMpc/$h$ yields $\tau_{\mathrm{eff}}$ measured across 50 cMpc/$h$ bins, the conventional value for Ly$\alpha$ forest opacity studies.

\begin{figure}
    \centering
    \includegraphics[width=\linewidth]{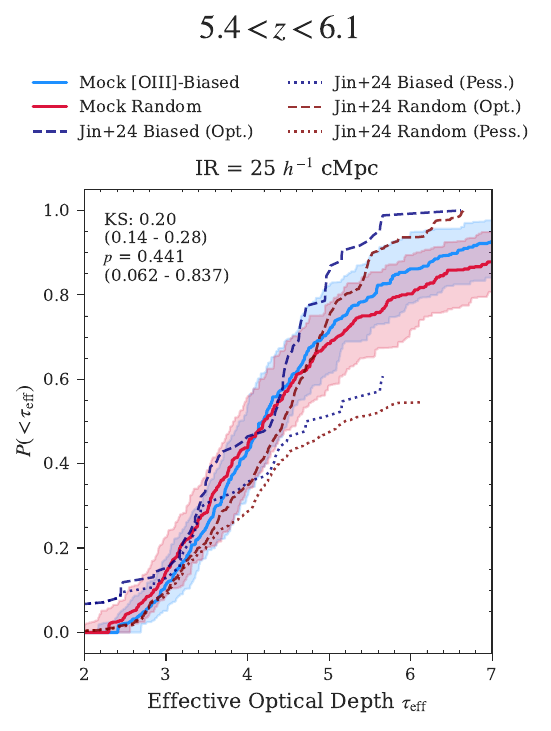}
    \caption{Cumulative distribution of two $\tau_{\text{eff}}$ distributions at $5.4 < z < 6.1$, for simulated [\ion{O}{iii}] emitters within an influence radius of 25 cMpc/$h$. The [\ion{O}{iii}]-biased distribution (blue) is built by computing $\tau_{\text{eff}}$ within bins determined by positions of nearby  mock [\ion{O}{iii}] emitters. The random distribution (red) is computed along random parts of the sightline, for $\tau_{\text{eff}}$ averaged over bins of the same length as for the biased distribution. We show the results of 100 realisations of our mock survey. The solid line shows the median of all surveys and the shaded bands represent the 16$^{\rm th}$ and 84$^{\rm th}$ percentile range ($\sim 1\sigma$). The corresponding optimistic (dashed) and pessimistic (dotted) CDFs presented in \protect\cite{Jin2024} are shown for comparison. We also indicate the median result of a Kolmogorov–Smirnov test and the median $p$-value after performing the test on each mock survey. In parentheses, we display the 16$^{\rm th}$ and 84$^{\rm th}$ range of these statistical tests to illustrate survey-to-survey scatter.}
    \label{CDF_Single}
\end{figure}

\begin{figure*}
    \centering
    \includegraphics[width=\textwidth]{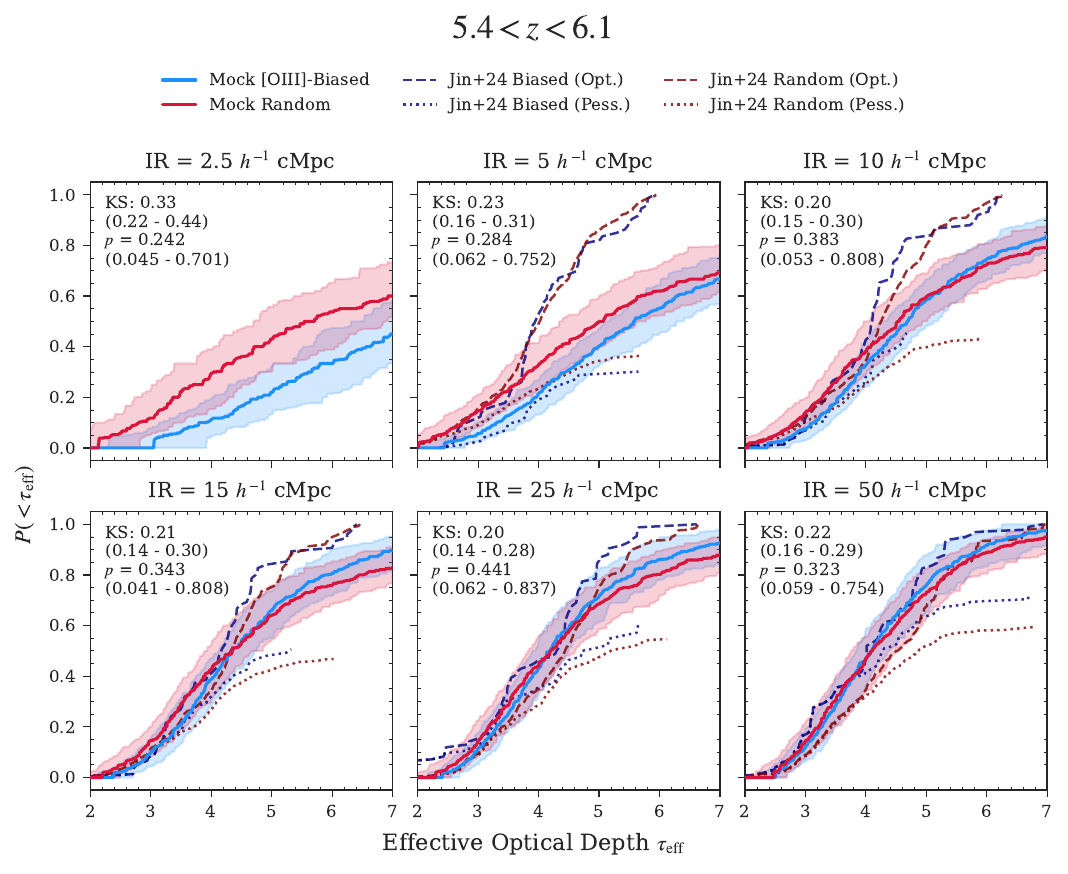}
    \caption{As for Fig. \ref{CDF_Single}, but with each panel showing a different influence radius, increasing from 2.5 to 50 cMpc/$h$.}
    \label{fig:CDF_1e11_z58}
\end{figure*}

\begin{figure*}
    \centering
    \includegraphics[width=\textwidth]{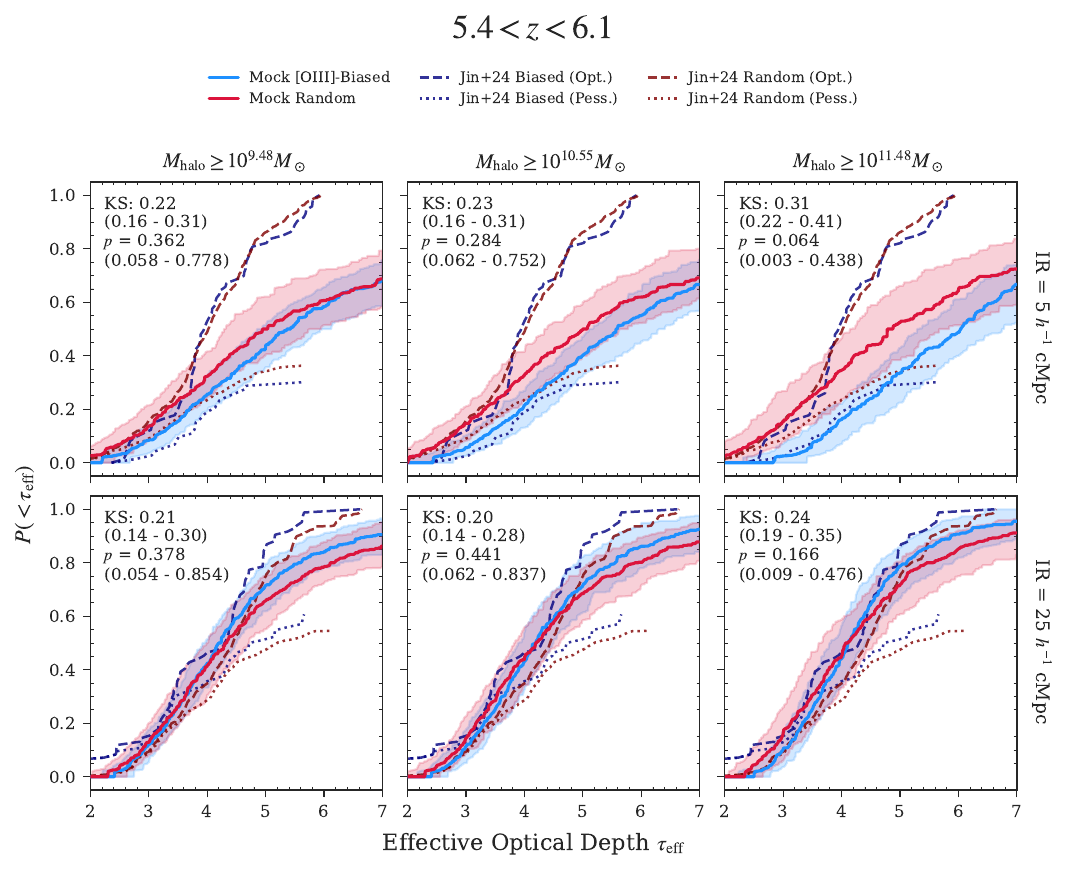}
    \caption{As for Fig. \ref{CDF_Single}, but exploring the effect of changing the minimum halo mass which is used to select the mock  [\ion{O}{iii}] emitters. The left column assumes a minimum mass $M_{\text{halo}}\geq10^{9.48}M_{\odot}$. The middle column shows our fiducial choice of $M_{\text{halo}}\geq10^{10.55}M_{\odot}$, motivated by JWST constraints on the clustering of [\ion{O}{iii}] emitters \citep{Huang2026, Eilers2024}. The right column assumes a minimum halo mass of $M_{\text{halo}}\geq10^{11.48}\;M_\odot$. We assume a different duty cycle in each case such that the number density of [\ion{O}{iii}] emitters is held constant. The top row shows results for an influence radius of 5 cMpc/$h$ and the bottom row assumes an influence radius of 25 cMpc/$h$.}
    \label{fig:CDF_masses}
\end{figure*}

\begin{figure*}
    \centering
    \includegraphics[width=\textwidth]{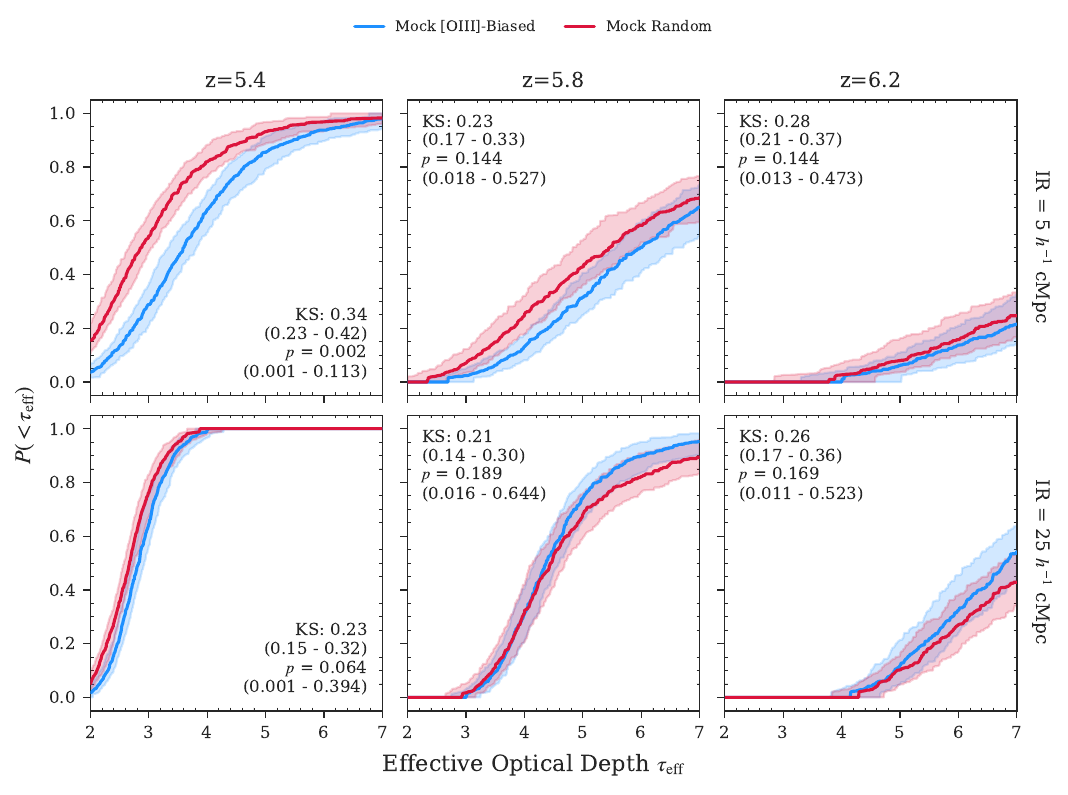}
    \caption{Similar to Fig. \ref{fig:CDF_masses}, except spectra are sampled at single redshift outputs. From the leftmost to the rightmost column, we present the $\tau_{\text{eff}}$ distributions for IR = 5 and 25 cMpc/$h$, near mock [\ion{O}{iii}] emitters at $z$ = 5.4, 5.8 and 6.2. These mock surveys are constructed to have a comparable total comoving distance as in \citet{Jin2024}, and realised 100 times. At high $z$ ($z=$ 5.8, 6.2), the large-scale IGM (IR = 25 cMpc/$h$) near [\ion{O}{iii}] emitters is more transparent than average. At low $z$ ($z=5.4$, 5.8), the local IGM (IR = 5 cMpc/$h$) is more opaque than average. }
    \label{fig:CDF_zs}
\end{figure*}

\begin{figure*}
    \centering
    \begin{minipage}[b]{0.48\textwidth}
        \centering
        \includegraphics[width=\linewidth]{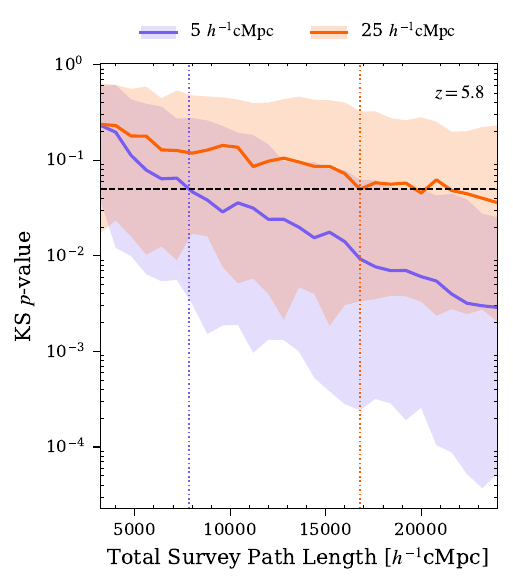}
    \end{minipage}
    \hfill
    \begin{minipage}[b]{0.48\textwidth}
        \centering
        \includegraphics[width=\linewidth]{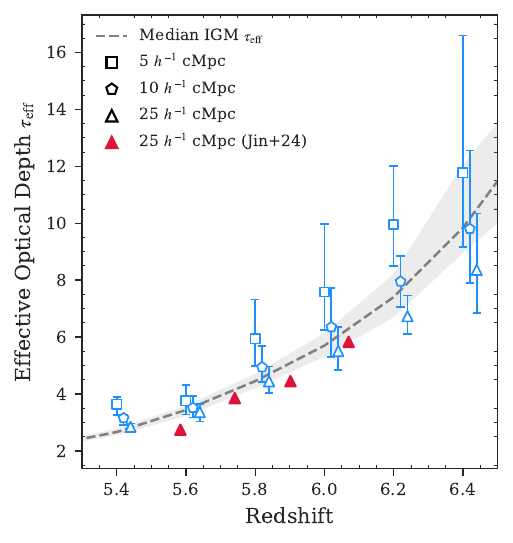}
    \end{minipage}
    \caption{\textbf{Left}: Effect of increasing the total comoving path length covered by an individual survey on the statistical significance of $\tau_{\mathrm{eff}}$ cumulative distributions. A survey is constructed using an integer number of 160 cMpc/$h$ sightlines at $z=5.8$ and the KS test is performed on the resultant biased and random $\tau_{\mathrm{eff}}$ CDFs. The purple curve illustrates median KS $p$-value convergence as a function of survey path length for $\tau_{\mathrm{eff}}$ distributions computed with IR = 5 cMpc/$h$. The orange curve illustrates the same effect for IR = 25 cMpc/$h$. The shaded bands represent the 16$^{\rm th}$ and 84$^{\rm th}$ percentile range ($\sim 1\sigma$) for 100 realisations of a given mock survey. The black horizontal dashed line indicates $p$ = 0.05. The vertical dotted lines mark where the median $p$-value for each influence radius crosses this threshold. At this point, we estimate the total number of probed sightlines is large enough on average to overcome cosmic variance. \textbf{Right}: The evolution of $\tau_\text{eff}$ near mock [\ion{O}{iii}] emitters as a function of redshift. The grey dashed line represents the median effective optical depth (computed over $50 \; \text{cMpc}/h$ bins) and its 1$\sigma$ scatter (grey shaded region). The blue markers represent the median $\tau_{\text{eff}}$ at increasing influence radii, with their error bars determined by the 1$\sigma$ scatter between different realisations of our mock surveys. Points from \protect\cite{Jin2024} have been included in red, for the case where the opacity was measured at an influence radius of 25 cMpc/$h$ from [\ion{O}{iii}] emitters. The $\tau_{\text{eff}}$ measurements are staggered by $\Delta z = 0.025$ for clarity. }
    \label{fig:tau_eff_vs_z}
\end{figure*}

\begin{figure*}
    \centering
    \includegraphics[width=\textwidth]{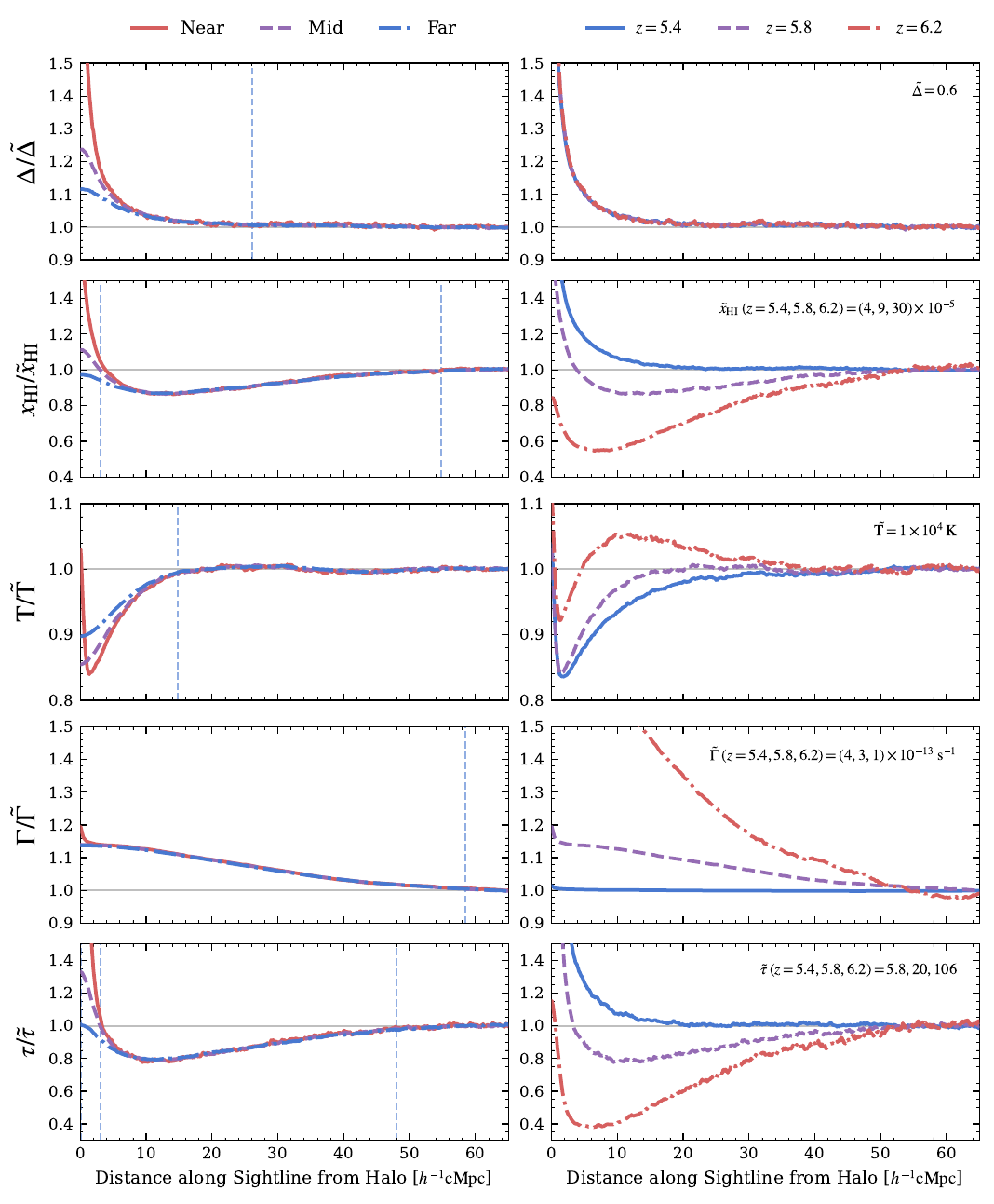}
    \caption{The evolution of IGM properties as a function of distance along a sightline close to a mock [\ion{O}{iii}] emitter. The rows, from top to bottom, display the gas density ($\Delta$), neutral hydrogen fraction ($x_{\mathrm{HI}}$), temperature ($T$), photoionisation rate ($\Gamma$) and Ly$\alpha$ optical depth ($\tau$). The first column shows stacked sightlines at $z=5.8$ where nearby haloes are located within various perpendicular distances to the origin of the sightline; near ($r_\perp\leq1\;\mathrm{cMpc}/h$), mid ($3\leq r_\perp\leq4\;\mathrm{cMpc}/h$) and far ($5\leq r_\perp\leq6\;\mathrm{cMpc}/h$). The properties are normalised by the median value at that redshift. Vertical dashed lines represent the radial distance at which the IGM is within $1\%$ of the background level. The second column presents how these quantities vary across different redshifts ($z=5.4,5.8$ and 6.2) for  mock [\ion{O}{iii}] emitters  near a sightline. Global median values are quoted for each redshift and in some cases this value is constant.}
    \label{fig:IGM_distances}
\end{figure*}

\section{Comparison with observations}
\label{sec:3}

Having constructed our mock survey, we next proceed to make a comparison with the observed Ly$\alpha$ forest opacity distributions measured in \citet{Jin2024}.

\subsection{Comparison at fixed influence radius}

In  Fig. \ref{CDF_Single}, we plot the $\tau_{\text{eff}}$ distributions measured close to galaxies and for a random sample. We show results for IR = 25 cMpc/$h$. The [\ion{O}{iii}]-biased sample is represented by the blue curve, where $\tau_{\text{eff}}$ measurements along a sightline are determined by the positions of nearby mock [\ion{O}{iii}] emitters. The bins over which $\tau_{\text{eff}}$ is averaged are centred at the position of the [\ion{O}{iii}] emitters. The bin size is determined by the distance between the source and the sightline and the assumed influence radius. The red curve represents the random sample, where $\tau_{\text{eff}}$ is measured at random points along a different sightline, using the same bin size as the biased sample. For an influence radius of 25 cMpc/$h$, the bin size for most $\tau_{\text{eff}}$ measurements is $\sim50 \mathrm{cMpc}/h$. The CDFs from \cite{Jin2024} are overlaid. The dashed curves represent the ``optimistic'' case, where we assume for regions where no transmitted flux is measured, the true flux is just below the $2\sigma$ detection limit (i.e., $\tau_{\mathrm{eff}} = \tau_{\mathrm{lim}, 2\sigma}$). For the ``pessimistic'' case, illustrated by the dotted line, the intrinsic transmitted flux is zero (i.e., $\tau_{\mathrm{eff}} = \infty$). These definitions follow the methodology introduced in \citet{Bosman2018}. 

Our mock CDFs are generally in good agreement with the observational data, although our simulation does not reproduce the line-of-sight chunks with the lowest Ly$\alpha$ forest opacity. These arise from observational measurements with a short path length, where the Ly$\alpha$ forest was truncated by the rest-frame 1040 \AA \,cut (see the right panel of Fig. 5 in \cite{Jin2024}). Our distribution of $\tau_{\text{eff}}$ near mock [\ion{O}{iii}] emitters displays a somewhat higher cumulative probability than the random sample for $\tau_{\mathrm{eff}}\gtrsim5$. This is similar to what is observed, implying that the effective optical depth of the IGM around ionising sources is lower on average compared to the average IGM. This arises naturally in our simulation, due to the inhomogeneous reionisation that we model. At lower opacities, we find that the distributions cross over, such that these are slightly more probable in the random distribution. We find significant variance in both of the distributions among our different realisations, as indicated by the shaded regions which show the 1$\sigma$ scatter.

We perform the two-sample Kolmogorov-Smirnov (KS) test for each of our 100 mock surveys in order to determine the likelihood that the two curves are drawn from the same underlying parent distribution. This is in contrast with \citet{Jin2024} who use a log-rank test to compare their CDFs, because their sample contains lower limits, so their distributions were first reconstructed using a Kaplan-Meier estimator. However, we found that for some individual surveys, the log-rank test ruled out the possibility that two samples were consistent, although their CDFs showed significant overlap. The reverse was also observed. We determined that this was because the log-rank test loses statistical power if the two distributions cross over regularly, as was commonly seen in our mock surveys. We therefore employ the KS test exclusively here.

The median KS test performed on each of our mock surveys returns a high $p$-value. The 1$\sigma$ interval also reveals considerable variance between survey repeats. Therefore, individual mock surveys lack the statistical power to capture these trends and a greater number of sightlines are necessary to reproduce the median distributions shown, suggesting that current surveys are dominated by cosmic variance. This is similar to what has been shown in modelling of the galaxy-Ly$\alpha$ forest cross-correlation \citep{Garaldi2024, Conaboy2026}.

\subsection{Effect of changing influence radius}
\label{sec:3.1}

We next explore the effect of changing the assumed influence radius. We note that this choice is independent of the physics driving reionisation in our simulation, as we are analysing only a single simulated source model. In Fig. \ref{fig:CDF_1e11_z58} we show a series of CDFs evaluated across influence radii ranging from 2.5 to 50 cMpc/$h$. Although observational data are currently unavailable for direct comparison at IR = 2.5 cMpc/$h$, we include this scale to theoretically explore the small-scale IGM. For the other influence radii plotted here, we also include the observational results of \citet{Jin2024}.

At the smallest influence radii (IR < 10 cMpc/$h$), we find the opacity of the Ly$\alpha$ forest is higher in the [\ion{O}{iii}]-biased sample than in the random sample. This is similar to the anti-correlation between galaxies and the IGM that is observed at small radii due to the bias of the Ly$\alpha$ forest \citep{Kakiichi2018}. The 1$\sigma$ scatter of our distributions is also large at these small influence radii, due to the variation in the environments of our mock [\ion{O}{iii}] emitters averaged across such small scales.

Moving to larger influence radii, we find that the distributions invert, with [\ion{O}{iii}]-biased regions becoming more transparent than the average IGM by IR = 25 cMpc/$h$. However, the difference between the biased and random CDFs in our simulation is not large, and there is significant overlap within the 1$\sigma$ scatter of our different realisations. Likewise, individual mock surveys return high $p$-values with significant variance. This implies that a single observational survey would require a substantially larger number of sightlines to reliably infer the median trends. Our median $p$-values at these intermediate influence radii are higher than those in \citet{Jin2024}. This may also reflect that the Sherwood-Relics simulations do not produce a peak in the galaxy-Ly$\alpha$ forest correlation at radii as large as observed (see \citealt{Conaboy2025} in comparison with \citealt{Kakiichi2025}).

As we reach the largest influence radius (IR = 50 cMpc/$h$), we find little distinction between the two samples. We note though that at an influence radius this large, the distance over which we are measuring $\tau_{\rm eff}$ is approximately 100 cMpc/$h$, close to our boxsize, which is still small in the context of simulations of reionisation \citep[compared with, e.g.,][]{zier2026}.

\subsection{Dependence on minimum halo mass}
\label{sec:3.2}

We further consider the impact of altering the minimum halo mass used to select our population of mock [\ion{O}{iii}] emitters. Fig. \ref{fig:CDF_masses} presents two alternative mass ranges and duty cycles, compared with the fiducial parameters which we take from \cite{Huang2026}. The leftmost column samples 0.094$\%$ of haloes with $M_{\text{halo}} \geq10^{9.48} M_\odot$ and the rightmost selects 100$\%$ of haloes with $M_{\text{halo}} \geq10^{11.48} M_\odot$. The middle column shows our fiducial mock [\ion{O}{iii}] sample. In each case, the resultant number density of  mock [\ion{O}{iii}] emitters  is 5.78 $\times 10^{-4} \;(\mathrm{cMpc}/h)^{-3}$, as we have tuned the duty cycle to ensure this remains constant. We show two influence radii for brevity; the first row probes short length scales near [\ion{O}{iii}] emitters (IR = 5 cMpc/$h$) and the second row probes large distances (IR = 25 cMpc/$h$). 

At IR = 5 cMpc/$h$, increasing the average mass of the mock [\ion{O}{iii}] emitter sample raises $\tau_{\text{eff}}$ in our biased sample, increasing the deviation between the biased and random distributions. This is due to the increasingly overdense regions traced by more massive haloes, which raises the overall opacity of the IGM, suggesting greater excess Ly$\alpha$ absorption near higher-mass haloes. At the larger value of IR = 25 cMpc/$h$, changing the halo mass has little effect on the difference between our biased and random samples. Ultimately, these distributions suggest that the qualitative trends of the galaxy-IGM connection remain unchanged, regardless of the mass of the underlying mock [\ion{O}{iii}] emitter population \citep[see also][]{Garaldi2024,Conaboy2026}. We interpret this as a consequence of source clustering within the influence spheres. A sphere centred on a high-mass halo likely also encompasses neighbouring lower-mass haloes, such that it becomes difficult to disentangle their individual effects when averaging the IGM over tens of Mpcs.

\subsection{Redshift evolution of the opacity CDFs}
\label{sec:3.3}

The final parameter we investigate is the evolution of the random and biased $\tau_{\rm eff}$ distributions with redshift. In Fig. \ref{fig:CDF_zs}, we show CDFs constructed from sightlines at three fixed redshifts ($z=5.4$, $z=5.8$ and $z=6.2$), as opposed to the CDFs shown previously which were constructed from lines-of-sight at a range of redshifts. We draw 22 sightlines through our volume at $z=5.4, 5.8$ and 6.2, close to the total path length probed by the entire ASPIRE Ly$\alpha$ forest over $z=5.4-6.1$, and generate 100 realisations of this as before. Each column compares two length scales, IR = 5 and 25 cMpc/$h$. For both of these choices of influence radius, we see that the median of the CDFs moves from lower to higher opacities with increasing redshift, and also becomes wider with increasing redshift, as expected \citep{Bosman2022}.

For the central column ($z=5.8$), we recover our previous trend where source-biased sightlines are more opaque than the random IGM for small IR, and slightly more transparent for large IR. This difference is somewhat more significant than in Fig. \ref{fig:CDF_1e11_z58}. The IGM evolves significantly over the range $5.4<z<6.1$, as reionisation finishes, and we find that strictly selecting sightlines from a fixed redshift range makes the difference between the random and biased samples clearer. This may motivate following up more quasar sightlines with JWST/NIRCam observations, to more tightly constrain local changes in the IGM opacity near galaxies. We also emphasise that, for this reason, it is important to try and match the redshift distribution of the simulation outputs and the observational data. 

At $z=6.2$, there is significant overlap in the distributions at both small and large IR, although the biased sample is somewhat more transparent (opaque) at large (small) influence radius. At $z=5.4$, the IGM of the random sample is more transparent for both large and small influence radii. The higher opacity of the IGM at small scales is in agreement with observations \citep[e.g.,][]{Adelberger2005, Turner2014}. Both of the distributions for IR = 25 cMpc/$h$ are narrow, implying a  $\tau_{\mathrm{eff}}$ with minimal sightline-to-sightline scatter. We attribute this to the rapid evolution in the mean free path of ionising photons at the end of reionisation \citep[e.g.,][]{Becker2021,Zhu_2023}, which dampens fluctuations in the UVB \citep{gaikwad2023, Davies2024}. After this, $\tau_{\mathrm{eff}}$ variance is strictly determined by the density and temperature field, such that the biased haloes show slightly lower Ly$\alpha$ transmission at large influence radii, as they are guaranteed to always pass through overdense regions.

A natural question to ask is then what observations would be required to mitigate the effect of cosmic variance. Fig. \ref{fig:tau_eff_vs_z} (left) explores the survey path length required to reach a $p$-value $<0.05$ between biased and random distributions for a given mock survey. We construct a test survey using an integer number of mock sightlines of length 160 cMpc/$h$ at $z=5.8$, and generate biased and random $\tau_{\mathrm{eff}}$ distributions for IR = 5 cMpc/$h$ (purple) and IR = 25 cMpc/$h$ (orange). We repeat each survey 100 times and compute the median $p$-value as well as the 1$\sigma$ percentile interval, represented by shaded bands. We increase the integer number of sightlines for each survey to determine the total absorption path length necessary to reduce the KS $p$-value $<0.05$, illustrated by the horizontal dashed line. The median survey path lengths required for each influence radius to reach significantly distinct distributions are marked with vertical dashed lines. We find that small influence radii (IR $=5$ cMpc/$h$) require roughly $8\times10^{3}$ cMpc/$h$ of Ly$\alpha$ forest to overcome cosmic variance and recover statistically distinguishable CDF curves using the KS test. This is a factor 4 increase in the existing path length probed by the ASPIRE survey. This increases to $(1.7\sim1.9)\times10^{4}$ cMpc/$h$ (or a factor 10 compared with ASPIRE) for large influence radii (IR $=25$ cMpc/$h$). The requirement further grows towards higher redshift as the IGM becomes more inhomogeneous, and samples constructed a range of redshift outputs from our simulation (as in Fig. \ref{fig:CDF_1e11_z58}), rather than being taken from a fixed simulation snapshot as in this estimate.

\subsection{Redshift evolution of the average IGM opacity}
 
Fig. \ref{fig:tau_eff_vs_z} (right) summarises the evolution of $\tau_{\text{eff}}$ at varying influence radii from mock [\ion{O}{iii}] emitters. We compare the median value of $\tau_{\text{eff}}$ at each redshift (grey dashed line and shaded region), measured from 100 mock surveys which each comprise 22 lines-of-sight (i.e. the total comoving distance covered by the Ly$\alpha$ forest in the ASPIRE survey). We further build $\tau_{\text{eff}}$ distributions for IR = 5, 10, and 25 cMpc/$h$ and take the median across 100 survey repeats. The error bars represent the 1$\sigma$ variance across repeats. 

The trends are consistent with those indicated by the CDFs: $\tau_{\text{eff}}$ near [\ion{O}{iii}] emitters is higher than the median IGM over small scales, and lower than the median over large scales. As redshift increases, the sightline-to-sightline scatter in $\tau_{\text{eff}}$ grows due to enhanced reionisation-related fluctuations. As the IR is increased, this scatter reduces because IGM inhomogeneities are smoothed out. The intrinsic scatter between sightlines is considerable at $z=6.4$, such that all $\tau_{\text{eff}}$ measurements are formally consistent with the average IGM even though the median points reveal an underlying trend. For $z=5.4$, we recover $\tau_{\text{eff}}$ trends observed in the low-$z$ regime, where we measure minimal scatter between surveys and no excess Ly$\alpha$ transmission at distances beyond IR = 10 cMpc/$h$. Excess IGM opacity for small IR increases between $z=5.6$ and $z=5.4$ as relic temperature fluctuations boost recombinations and grow the \ion{H}{i} population within source overdensities. We also plot the $\tau_{\text{eff}}$ measurements near ASPIRE [\ion{O}{iii}] emitters at IR = 25 cMpc/$h$ \citep{Jin2024}, with which we find close agreement.

\section{Connection to IGM properties}
\label{sec:4}

The large scatter in $\tau_{\mathrm{eff}}$ beyond $z\sim5.5$ \citep{Becker2015, Bosman2018, Eilers2018} cannot be accounted for by fluctuations in the density field alone \citep{Becker2015}. Instead, this scatter must be considered in the context of fluctuations in the UV background \citep{Davies2016,daloisio2018}, variations in temperature due to an extended and inhomogeneous reionisation \citep{DAloisio2015, keating2018} and islands of residual neutral hydrogen below $z=6$ \citep{Kulkarni2019, keating2020, nasir2020}. Therefore, we are motivated to explore the evolution of these quantities near [\ion{O}{iii}] emitters, to determine the physical mechanisms responsible for the observed changes in $\tau_{\mathrm{eff}}$.

\subsection{Evolution with increasing perpendicular distance}

The left column of Fig. \ref{fig:IGM_distances} shows the properties of the IGM along a quasar sightline at increasingly large line-of-sight and perpendicular distances from a mock [\ion{O}{iii}] emitter. To characterise this, we take random lines of sight through our simulation volume, and select those that pass close to all haloes of mass $M_{\mathrm{halo}}>10^{10.55}M_\odot$. We measure the perpendicular distance between the halo and its closest point to the sightline and group these into three bins of varying perpendicular distance: ``near'' ($r_\perp\leq1\;\mathrm{cMpc}/h$), ``mid'' ($3\leq r_\perp\leq4\;\mathrm{cMpc}/h$) and ``far'' ($5\leq r_\perp\leq6\;\mathrm{cMpc}/h$). We shift these sightlines such that the closest point between the halo and line-of-sight lies at the midpoint of the sightline, in order to stack the sightlines about a common origin. The following gas properties are available along each sightline: gas density ($\Delta$), neutral hydrogen fraction ($x_{\mathrm{HI}}$), temperature ($T$), ionisation rate ($\Gamma$) and Ly$\alpha$ optical depth ($\tau$). We plot the median value of each quantity across all stacks, normalised by the global median for a given redshift. In this way, we construct a profile that illustrates how the median IGM varies along a quasar sightline with an [\ion{O}{iii}] emitter roughly positioned at $r=0$. Although we solely show the median here for clarity, we emphasise that the sightline-to-sightline variance is substantial. For the left-hand column, all sightlines are taken from a simulation snapshot at $z=5.8$. We include vertical dashed lines to indicate the radial distances at which the profiles are within 1$\%$ of the median value in our simulation.  
We observe the same trends for our three different bins of halo perpendicular distance. At small line-of-sight distances ($r\lesssim 5\;\mathrm{cMpc}/h$), the IGM is typically overdense and maintains a higher-than-average $x_{\mathrm{HI}}$, driving a high optical depth. The IGM is cooler than average here, as these regions were ionised first and have since had time to cool via adiabatic expansion, inverse Compton scattering and collisional excitation. At greater line-of-sight distances ($r\gtrsim5\;\mathrm{cMpc}/h$), the density reduces more rapidly than the photoionisation rate, leading to an intermediate scale where the gas density is close to background levels, but the photon density remains preferentially higher than average. Here, $x_{\mathrm{HI}}$ is lower and the IGM is more transparent than average. This extends out to line-of-sight distances approaching $\sim50\;\mathrm{cMpc}/h$, at which point the ionisation rate reduces enough such that the IGM resembles the background. Probing  mock [\ion{O}{iii}] emitters  at increasing perpendicular radii from the sightline has little overall effect on the shape of the profiles. Variance is most prominent at short perpendicular distances, particularly when the line-of-sight distance along the sightline is less than the perpendicular distance to the source. In this case, the points of closest approach for haloes ``near'' a sightline (within $1\;\mathrm{cMpc}/h$) correspond to the strongest density peak, maximal temperatures and the highest $x_{\mathrm{HI}}$ and $\tau$.

\subsection{Evolution with redshift}

The next column illustrates how these properties change across different snapshots ($z$ = 5.4, 5.8 and 6.2). For these profiles, we strictly use haloes within a perpendicular distance of 1 cMpc/$h$  from the closest point to the sightline. The average density field changes a negligible amount between $5.4<z<6.2$. Instead, changes in the IGM opacity at larger line-of-sight radii are driven by redshift evolution in the photoionisation rate, temperature and residual neutral islands. We find that the contrast between local and average IGM properties increases with increasing redshift, in tandem with the evolution of the mean free path of ionising photons from sources in our simulation \citep{Feron24}.

At $z=6.2$, the local ionisation field of  mock [\ion{O}{iii}] emitters  is roughly twice the global background. Close to haloes, the IGM is more ionised, more transparent and cooler than average, indicating these regions were ionised first. Escaping photons are heating and ionising regions at intermediate distances from the central overdensity. At its maximum ($r\sim6\; \mathrm{cMpc}/h$), $\tau$ is at 40$\%$ of the background level and $x_{\mathrm{HI}}$ is close to half of the median value. These effects are prominent until line-of-sight distances close to $\sim50\;\mathrm{cMpc}/h$ from the source, beyond which these quantities reach background levels. During this earlier stage of reionisation, we expect the amplitude of excess transmission to increase due to the increase in temperature near haloes. The global minimum for $x_{\mathrm{HI}}$ occurs at a larger radial distance at $z=5.8$ than $z=6.2$. This trend supports the consensus of an inside-out reionisation picture, where overdensities were likely ionised first and ionised bubbles expanded radially outwards from haloes.

At $z=5.8$, the IGM close to the  mock [\ion{O}{iii}] emitters  has cooled further, supporting recombination and a higher-than-median $x_{\mathrm{HI}}$ within small line-of-sight distances ($r\lesssim 5\;\mathrm{cMpc}/h$). As the global ionising background rises, the contrast between the global and local photoionisation rate decreases. Similarly, the global median $x_{\mathrm{HI}}$ increases as the remaining neutral voids start to become ionised. Intermediate line-of-sight distances from haloes ($r\sim10\; \mathrm{cMpc}/h$) remain more ionised than average, but the difference from the background value is smaller. At these distances, the temperature resembles that of the median IGM.

By $z=5.4$, ionising photons have dispersed evenly throughout the IGM and the ionisation field becomes spatially homogeneous. Therefore, the changing IGM optical depth is strictly determined by the effects of the density and temperature field. Regions closest to the mock [\ion{O}{iii}] emitters are both overdense and cool, enhancing the recombination rate and resulting in higher $x_{\mathrm{HI}}$ and a higher IGM opacity. Beyond a line-of-sight distance of 10 cMpc/$h$, the median optical depth along our biased sightlines is identical to the background opacity.

\section{Conclusions}
\label{sec:5}

In this paper, we have performed the first theoretical investigation of how the distribution of IGM opacities near galaxies compares to a random sample at the end of reionisation. We investigated the Ly$\alpha$ forest using a patchy reionisation simulation from the Sherwood-Relics simulation suite, to determine the influence of mock [\ion{O}{iii}] emitters on the effective optical depth of their surrounding IGM, and compared this to observations from the ASPIRE survey \citep{Jin2024}. We modelled a population of [\ion{O}{iii}] emitters and correlated them with a combination of mock Ly$\alpha$ forest spectra between $5.4<z<6.1$, taking into account the evolution of the IGM along this redshift range as well as the survey geometry enforced by the JWST/NIRCam field-of-view. We constructed a ``biased'' distribution of Ly$\alpha$ forest opacities, which measured the mean transmitted flux close to [\ion{O}{iii}] emitters, and contrasted this with a random sample. Our main results are as follows:

\begin{itemize}
    \item In agreement with previous works analysing the galaxy-Ly$\alpha$ forest cross-correlation both after and during reionisation, we found that at small influence radii, our biased sample showed higher opacity than the random sample, due to the overdense regions probed by these haloes. 
    \item At larger influence radii, we tentatively found that regions of the IGM with a lower-than-average opacity were more likely to be observed in our biased sample, due to a local enhancement of the photoionisation rate around the mock [\ion{O}{iii}] emitters.   However, we also found that regions with higher-than-average opacity were also more likely to be observed. As well as this, within different realisations of our mock surveys, we found significant overlap between the biased and random samples, and the results were not statistically significant according to a KS test.
    \item We found that these results were insensitive to how we constructed our mock [\ion{O}{iii}] emitter catalogues, with our distributions showing similar results when our [\ion{O}{iii}] emitters were placed in higher or lower halo masses, as long as the total number density was held constant.
    \item The biased and random opacity distributions evolved significantly with redshift. Although the biased region showed higher opacity at small scales independent of redshift, at larger scales the difference between the random and biased samples increased with increasing redshift.
\end{itemize}

In general, we found reasonable agreement between the Sherwood-Relics simulation and the ASPIRE survey. As pointed out in \cite{Garaldi2024}, the degeneracy between the contributions of high-mass and fainter neighbouring haloes cannot be fully resolved using this geometric approach, meaning that source clustering likely plays an important role in boosting the apparent ionisation bias near [\ion{O}{III}] emitters. Within different realisations of our mock surveys, we additionally found significant overlap between the biased and random samples. Increasing the number of sightlines considered within a given redshift bin helped to discriminate between two samples; we estimate a survey like ASPIRE, but a factor 4 larger in Ly$\alpha$ forest path length is necessary for small-scale studies and at least a factor 10 times larger for larger influence radii. This suggests that a larger observational sample of galaxies neighbouring quasar sightlines will be important for a more accurate characterisation of the impact of galaxies on their local IGM, either through cross-correlating the Ly$\alpha$ forest along more quasar sightlines with new galaxies identified in future JWST observations or looking ahead to new ground-based surveys with VLT/MOONS \citep{moons2020} and future instruments like ELT/MOSAIC \citep{mosaic2021}. At the same time, new reionisation simulations in larger volumes exploring different reionisation histories and source models will be crucial for quantifying the modelling uncertainty. New observational data together with these more realistic models will guide us towards more accurate representations of the processes that shaped large-scale structure, and ultimately, enhance our understanding of how cosmic environments evolved to form the Universe as we observe it today.

\section*{Acknowledgements}

LCK acknowledges the support of a Royal Society University Research Fellowship (grant number URF$\backslash$R1$\backslash$251793). LC and JSB are supported by STFC consolidated grant ST/X0009821/1. We thank Volker Springel for making \textsc{p-gadget-3} available. We also thank Dominique Aubert for sharing the \textsc{aton} code. The simulations used in this work were performed using the Joliot Curie supercomputer at the Tr\'{e}s Grand Centre de Calcul (TGCC) and the Cambridge Service for Data Driven Discovery (CSD3), part of which is operated by the
University of Cambridge Research Computing on behalf of the STFC DiRAC HPC Facility (www.dirac.ac.uk). We acknowledge the Partnership for Advanced Computing in Europe (PRACE) for awarding us time on Joliot Curie in the 16th call. The DiRAC component of CSD3 was funded by BEIS capital funding via STFC capital grants ST/P002307/1 and ST/R002452/1 and STFC operations grant ST/R00689X/1. This work also used the DiRAC@Durham facility managed by the Institute for Computational Cosmology on behalf of the STFC DiRAC HPC Facility. The equipment was funded by BEIS capital funding via STFC capital grants ST/P002293/1 and ST/R002371/1, Durham University and STFC operations grant ST/R000832/1. DiRAC is part of the National e-Infrastructure. We acknowledge the use of OpenAI's ChatGPT language model, which assisted in debugging and refining parts of the analysis and visualisation code used in this work.

\section*{Data Availability}

All data and analysis code used in this work are
available from the first author on reasonable request.
Further guidance on accessing the publicly available Sherwood-Relics simulation data may also be
found at \url{https://www.nottingham.ac.uk/astronomy/sherwood-relics/}.



\bibliographystyle{mnras}
\bibliography{refs} 



\bsp	
\label{lastpage}
\end{document}